# Calcul haute performance et efficacité énergétique : focus sur OpenFOAM

**Cyrille Bonamy**
LEGI – UMR5519 – CNRS – EcoInfo
1209-1211 rue de la piscine
Domaine Universitaire
38400 Saint Martin d'Hères

**Laurent Lefèvre**
LIP – UMR5668 – Inria – EcoInfo
Ecole Normale Supérieure de Lyon
46, allée d'Italie
69364 Lyon Cedex 07

**Gabriel Moreau**
LEGI – UMR5519 – CNRS – EcoInfo
1209-1211 rue de la piscine
Domaine Universitaire
38400 Saint Martin d'Hères

## Résumé

*Le calcul haute performance est de plus en plus utilisé au sein de la société. Auparavant réservé à une élite, basé sur des grandes infrastructures de calcul et stockage, c'est désormais une brique de base de bon nombre d'entreprises. En effet, le calcul haute performance permet de concevoir et d'optimiser de nombreux éléments pour un coût limité comparativement à la réalisation de prototypes ou d'essais in situ. Il est également fortement utilisé dans tout ce qui concerne le big data et l'intelligence artificielle.*

*Il apparaît primordial de se poser la question de l'impact environnemental de ces pratiques numériques. Un certain nombre d'actions a déjà été initié dans cette communauté : GREEN500 [1] ; label européen d'éco-responsabilité CoC [2] pour les datacentres ; formation au sein du GDS EcoInfo [3], prise en compte d'utilisation de leviers de réduction énergétique dans les centres de calculs [4]… mais certaines de ces actions considèrent des situations et/ou des logiciels spécifiques voire idéalisés.*

*La démarche de qualification d'un logiciel dans le domaine du calcul haute performance consiste à regarder la scalabilité du logiciel. L'originalité de cette étude consiste à se focaliser sur la scalabilité énergétique (temps de retour du calcul en fonction de la puissance consommée), en considérant plusieurs architectures (trois machines du TOP500 et un cluster de laboratoire).*

*Le coût énergétique d'un calcul exemple a pu être estimé, il en ressort que la machine la plus efficace en termes de temps de calcul n'est pas forcément la plus efficace énergétiquement, que selon le nombre de cœurs/processus choisi, ce n'est pas toujours*

*la même architecture la plus rentable énergétiquement. Il a ainsi été possible de mettre en évidence que plus l'utilisateur est prêt à attendre, moins le calcul coûte énergétiquement.*

*Il est par ailleurs important de sensibiliser les utilisateurs de moyens de calcul au fait que l'impact environnemental d'un calcul dépasse le coût énergétique direct des calculs. En effet, pour exécuter de tels calculs, les lourdes infrastructures nécessaires (nœuds de calcul, équipements réseaux, stockage…) ont un cycle de vie qui génère de multiples impacts environnementaux : pour l'extraction des ressources (métaux et terres rares), leur fabrication, transport et la fin de vie (difficulté du recyclage).*

## Mots-clefs

*Impact environnemental, HPC, OpenFOAM, Green500*

## 1 Contexte

Le calcul haute performance (HPC) est de plus en plus utilisé au sein de la société. Auparavant réservé à une élite (communauté scientifique, multinationale), le calcul scientifique est désormais une brique de base de bon nombre d'entreprises, qu'elles soient grosses ou petites (PME, jeune pousse…). En effet, le calcul haute performance permet de nos jours de concevoir et d'optimiser de nombreux éléments pour un coût limité comparativement à la réalisation de prototypes ou d'essais *in situ*. Il est également fortement utilisé dans tout ce qui concerne le Big data ou encore l'Intelligence Artificielle.

Il apparaît ainsi important de se poser la question de l'impact environnemental de ces pratiques numériques.

Dans cet article, nous proposons de regarder l'impact environnemental du calcul haute performance, afin de sensibiliser et d'alerter la communauté calcul sur ce sujet. Un certain nombre d'actions ont déjà été initiées dans cette communauté : mise en place du GREEN500 [1] ; label européen d'éco-responsabilité CoC [2] pour les datacentres ; formations au sein du GDS EcoInfo [3], prise en compte d'utilisation de leviers de réduction énergétique dans les centres de calculs [4]… mais certains de ces travaux considèrent des situations et/ou des logiciels idéalisés (type test Linpack [5]).

Notre étude est basée sur un cas concret réel récent et permet de voir l'efficacité énergétique du code sur plusieurs architectures en fonction du nombre de cœurs/processeurs considérés.

## 2 Démarche et méthodologie

### 2.1 Présentation générale de la démarche

La démarche classique de qualification d'un logiciel dans le domaine du calcul haute performance est de regarder la scalabilité forte et/ou faible du logiciel. Celle-ci est généralement représentée via le *speed-up* du calcul, à savoir le ratio du temps de calcul en séquentiel (sur un seul cœur) sur le temps de calcul en parallèle sur N cœurs. Ici, nous considérons uniquement la scalabilité forte du code. Il s'agit pour un calcul identique avec un nombre constant de cellules de faire varier le nombre de cœurs et donc le découpage du maillage. La grandeur d'intérêt de cette étude est l'énergie consommée par ce même calcul.

Le cas pratique considéré est issu de la thèse de Tim Nagel [6] effectuée au laboratoire LEGI (Laboratoire des Écoulements Géophysiques et Industriels), sous la direction de Achim Wirth, Julien Chauchat et Cyrille Bonamy. L'objectif de ce doctorat était d'étudier le phénomène d'érosion d'un écoulement autour d'un cylindre vertical. C'est un problème difficile de simulation numérique portant sur la résolution des équations de la mécanique des fluides. Pour ce faire, on utilise le solveur diphasique opensource libre sedFoam [7] développé en interne au laboratoire LEGI en étroite collaboration avec l'Université du Delaware. Ce solveur est basé sur le logiciel libre OpenFOAM [8] qui est une brique logicielle de plus en plus utilisée dans la communauté de la mécanique des fluides, que ce soit en recherche ou dans les entreprises tant en recherche que dans les entreprises. À noter que toute la partie parallélisation du code est ainsi faite au travers de la bibliothèque Pstream qui est une surcouche à la librairie classique MPI.

Lors de cette étude de la scalabilité énergétique, un seul et même maillage de six millions de cellules a été utilisé et une seule seconde de la dynamique de l'écoulement a été simulé. En temps réel sur supercalculateur, cela correspond entre une à huit heures de calcul en fonction du nombre de cœurs considérés et de l'architecture utilisée. À noter que lors de la thèse, six cents secondes de dynamique ont été simulées mais sur cet exemple, le temps pris pour chaque seconde est relativement constant d'une seconde à l'autre.

Les architectures informatiques testées sont :

- la machine Occigen du CINES/GENCI [9] (77e au TOP500 et 142e au GREEN500 : bullx DLC, Xeon E5-2690v3 2x14C 2.6 GHz, Infiniband FDR) ;
- la machine JOLIOT-CURIE KNL du CEA/TGCC-GENCI (242e au TOP500 et 42e au GREEN500 : Bull Sequana X1000, Intel Xeon Phi 7250 68C 1.4 GHz, Bull BXI 1.2) ;
- la machine JOLIOT-CURIE SKL du CEA/TGCC-GENCI (40e au TOP500 et 34e au GREEN500 : Bull Sequana X1000, Xeon Platinum 8168 2x24C 2.7 GHz, Mellanox EDR) ;
- les moyens locaux du laboratoire LEGI (un cluster composé de nœuds DELL PowerEdge C6320 2x10C 2.4 GHz, interconnectés via un réseau 10 Gb/s – non classé !).

## 2.2 Méthodologie

### 2.2.1 Mesure de la scalabilité forte

La première chose qui nous est apparue importante est de vérifier la bonne scalabilité d'OpenFOAM, et de définir celle-ci.

Comme expliqué précédemment, nous avons limité notre étude à la scalabilité forte, c'est-à-dire à reproduire un calcul identique (même maillage de six millions de cellules, une seconde de la dynamique de l'écoulement simulé) sur un nombre de cœurs de calcul N variable. La scalabilité dite faible, qui revient à faire évoluer la taille du maillage conjointement avec le nombre de cœurs de calcul N n'a pas été considérée ici.

Afin de mesurer la scalabilité forte, un calcul d'une seconde de simulation physique a été réalisé sur un certain nombre de nœuds. Les nœuds ont toujours été chargés au maximum, c'est-à-dire que nous avons utilisé tous les cœurs de chaque nœud considéré, ce qui se fait classiquement dans le monde du calcul scientifique. Par exemple, sur l'infrastructure SKL du TGCC, nous avons considéré un multiple de 24 cœurs.

Afin de représenter au mieux cette scalabilité, le temps de calcul a été dimensionné par un temps référence choisi *a posteriori* parmi les plus efficaces. Ce temps de référence considéré est le temps du calcul sur un nœud CINES/Occigen (28 cœurs/processus MPI) ramené à un cœur, de manière à pouvoir comparer les divers points à une courbe de scalabilité idéale qui est défini par 1/N, N étant le nombre de cœurs/processus MPI.

La figure 1 ci-dessous rend compte des caractéristiques des calculs réalisés lors de ce travail.

### 2.2.2 Consommation électrique

La consommation électrique a été estimée pour l'ensemble des calculs précités.

Seuls les calculs sur le cluster local du laboratoire LEGI sont associés à une vraie mesure de la consommation électrique. Ces mesures sont effectuées au travers de commandes SNMP exécutées au début et à la fin des calculs et qui permettent d'interroger les « multiprises » PDU monitorées auxquelles sont branchés les serveurs.

Concernant les calculs sur les centres de calcul nationaux (CEA et CINES), les données issues du référentiel Green500 [1] (consommation instantanée en kW pour un certain nombre de cœurs) ont été utilisées afin d'estimer à partir du temps de calcul la consommation électrique. Une simple règle de trois a ainsi été appliquée.

Cette méthodologie, triviale permettant une estimation rapide, est critiquable. Mais nos analyses (non présentées ici) des mesures effectuées au laboratoire LEGI nous permettent d'affirmer que cette estimation est représentative de la réalité.

## 3 Résultats

Ces travaux ont permis de confirmer la bonne scalabilité forte d'OpenFOAM (ainsi que la limite habituelle de ce type de code : un minimum de 30 000 cellules par processus MPI est nécessaire afin d'être efficace ; voir la figure 1).

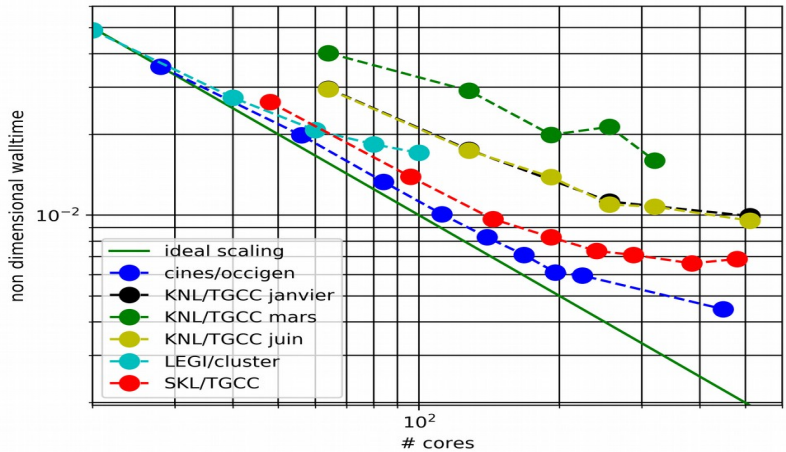

*Figure 1 - Scalabilité forte*

Le coût énergétique des calculs considérés dans cette étude a pu être estimé. Les résultats, présentés sur la figure 2, montrent que la machine la plus efficace en termes de temps de calcul n'est pas forcément la plus efficace énergétiquement.

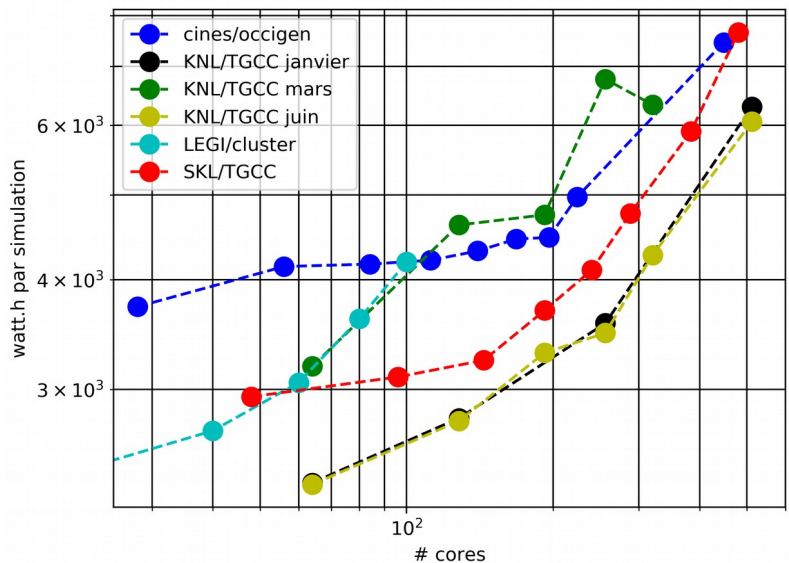

*Figure 2 - Scalabilité énergétique*

Une représentation rare de la scalabilité est également réalisée : le temps de retour d'un calcul versus le coût énergétique de ce même calcul (figure 3). Il permet de se rendre compte que « **plus on est prêt à attendre, moins le calcul coûte énergétiquement** ». Mais il montre aussi que ce n'est pas toujours la même architecture qui est la plus rentable énergétiquement. En effet, on observe que les courbes se croisent sur cette figure 3.

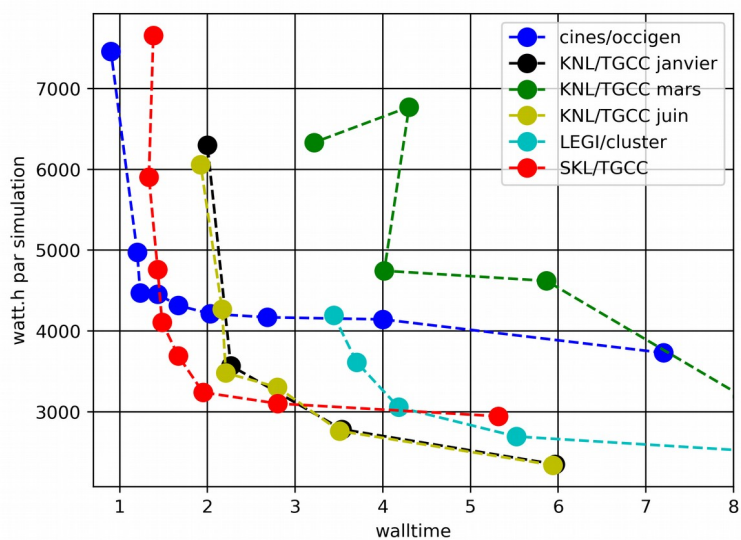

*Figure 3 - Coût énergétique versus temps de retour*

Un point intéressant non prévu est également ressorti de ce travail. Entre janvier et mars 2019, il y a eu une importante régression de performances sur la machine KNL du CEA/TGCC (cf. figure 1, différence de 30 à 40 % entre janvier et mars). De nombreux échanges avec le CEA ont eu lieu, mais il n'a pas été possible d'identifier la raison de cette évolution des performances. Plusieurs pistes ont été évoquées (évolutions concernant la couche MPI, mises à jour de micro-codes matériels). Lorsque l'on a repris le travail en juin 2019, les performances de janvier ont été retrouvées. Même si aucune conclusion franche ne peut être amenée, ce résultat montre la complexité des infrastructures actuelles, ainsi que l'importance du logiciel, des librairies associées, des micro-codes et de leurs paramétrages. Ces points sont rarement considérés actuellement.

## 4 Conclusion

Nos résultats nous permettent de proposer en guise de conclusion **quelques bonnes pratiques à suivre**.

Tout d'abord, on ne peut que conseiller aux utilisateurs de moyens de calcul parallèle d'étudier la scalabilité de leurs calculs (point qui est déjà quasi imposé par les centres de calcul nationaux gérés par GENCI), et ainsi de faire en sorte de limiter leur impact. Dans bon nombre de situations, le besoin d'un résultat « immédiat » n'est pas justifié ; le calcul peut ainsi être exécuté sur un nombre limité de cœurs de calcul. Ainsi, quelques heures ou jours de patience peuvent permettre une économie substantielle. Un autre axe permettant de limiter le temps de calcul (et ainsi la consommation énergétique) consiste à limiter au minimum le nombre de cellules.

La seconde bonne pratique concerne le choix de l'infrastructure. Quand cela est possible, il peut être intéressant de changer d'infrastructure afin de réduire la consommation induite par ses calculs. Pour ce faire, une étude préalable des architectures disponibles est à accomplir afin de déterminer, en fonction du cas d'usage, la plus efficace.

Une troisième bonne pratique qui peut paraître évidente, mais qui n'est pas toujours considérée, est la consommation inhérente à ce type d'infrastructures : une machine qui ne fait rien consomme énormément, et sa durée de vie n'en est pas améliorée. Un remplissage optimal des infrastructures est donc primordial pour une efficacité globale optimum avec si possible une mise à l'arrêt automatique des nœuds non utilisés, puis un redémarrage lui aussi sans intervention en cas de besoin. Le scheduleur libre OAR [10] développé sur Grenoble propose cette fonctionnalité.

De plus, même si ce point n'a pas été abordé dans cette étude, il a été montré lors de la formation ANF EcoInfo « comprendre et agir » [11], que réduire les sorties (logs, visualisations) peut permettre un gain important.

Les évolutions prévisibles des machines d'ici quelques mois / années devraient aussi permettre d'améliorer encore l'efficacité au sens énergétique du calcul haute performance. En effet, l'avènement de l'Intelligence Artificielle entraîne l'essor du calcul sur carte graphique, ou encore sur co-processeur Intel, sur processeurs ARM, voire sur des plateformes plus spécifiques comme les Nvidia Jetson. De nombreux logiciels peuvent *a priori* profiter de ces matériels. Il est cependant de plus en plus compliqué de porter les codes de calcul, eux-mêmes de plus en plus complexes, sur ces architectures. La boîte à outils OpenFOAM est d'ailleurs un bon exemple de ce dernier point. Elle constitue pour la majorité de ses utilisateurs un langage de haut-niveau. Il est donc difficile d'envisager un portage de son code sur une architecture novatrice. De nombreux essais ont été réalisés dans ce sens, mais ils sont à ce jour restés infructueux.

La communauté scientifique peut cependant espérer un effort des constructeurs afin de rendre automatique le portage. La nouvelle version de la bibliothèque d'OpenMP [12] va dans ce sens. Tout comme OpenMPI est devenu un standard, on peut espérer qu'un standard émerge des bibliothèques actuelles et permette le portage de codes sur ces nouvelles architectures.

Cette conclusion ne peut se finir ainsi. En effet, toute l'étude présentée ne concerne que le coût énergétique direct des calculs, c'est-à-dire la phase d'usage. L'impact environnemental d'un calcul dépasse largement cette phase d'usage. Il faudrait considérer l'ensemble du cycle de vie du service numérique associé. Pour être plus concret, le service numérique à considérer ici revient à répondre à l'Unité Fonctionnelle suivante : « calculer une seconde de dynamique du problème physique considéré (érosion autour d'une pile de pont) ». Ce service englobe donc à la fois les aspects développement/usage (choix du logiciel, choix des paramètres numériques, maillage…), ainsi que les aspects infrastructure (choix de l'infrastructure : nœuds de calcul, équipements réseaux, stockage, climatisation…). Pour analyser de manière globale l'impact environnemental du service numérique, il faut ainsi prendre en compte le reste de l'ensemble du cycle de vie du matériel informatique utilisé : l'extraction minières des ressources (métaux et terres rares…), la fabrication, le transport et la fin de vie (complexité du recyclage et de la récupération des métaux).

On comprend ainsi très vite la complexité d'une telle étude et des éventuelles comparaisons entre moyens de répondre à un « service numérique ». Il est très difficile de dire si telle ou telle solution est « meilleure ». En fonction des critères et hypothèses considérés, on peut dire tout et son contraire.

Au travers de ce papier et de la présentation orale associée, nous espérons avoir atteint notre objectif premier : sensibiliser les utilisateurs et les administrateurs systèmes et réseaux de l'importance du logiciel et de l'architecture matérielle sous-jacente à l'impact environnemental global de leurs calculs hautes performances.

## Remerciements

## Bibliographie